\documentclass{article}
\usepackage{spconf,amsmath}
\usepackage{graphicx,color,xcolor}
\usepackage{mathtools}
\usepackage{amsthm, amssymb,latexsym}
\usepackage{amsthm}
\usepackage{color}
\usepackage{subfigure}
\usepackage{comment}
\usepackage{hyperref}
\usepackage{booktabs} 

\def\wtilde{\widetilde}

\def \d{\delta}

\def \eps{\epsilon}

\title{On the robustness of non-intrusive speech quality model by adversarial examples}

\name{Hsin-Yi Lin, Huan-Hsin Tseng, Yu Tsao}
\address{Research Center for Information Technology Innovation, Academia Sinica, Taipei, Taiwan}

\begin{document}

\maketitle
\begin{abstract}
It has been shown recently that deep learning based models are effective on speech quality prediction and could outperform traditional metrics in various perspectives. Although network models have potential to be a surrogate for complex human hearing perception, they may contain instabilities in predictions. This work shows that deep speech quality predictors can be vulnerable to adversarial perturbations, where the prediction can be changed drastically by unnoticeable perturbations as small as $-30$ dB compared with speech inputs. In addition to exposing the vulnerability of deep speech quality predictors, we further explore and confirm the viability of adversarial training for strengthening robustness of models.
\end{abstract}





\section{Introduction}
\label{sec:intro}

The need for speech quality evaluation has been raised as the increasing use of speech processing algorithms and telecommunications applications. Traditionally, the Mean Opinion Score (MOS) of a speech sample is derived from averaging subjective listening tests~\cite{rec1996p,rec2018p, itu2003835} of participants. Although actual human rating is considered the most faithful index to assess speech quality, listening tests are typically costly and time consuming. To reduce the last two factors, automatic speech quality predictions mimicking human perceptions have become an active research topic. Several metrics such as Perceptual Evaluation of Speech Quality (PESQ) \cite{recommendation2001perceptual} and  Short-Time Objective Intelligibility (STOI) \cite{taal2010short} were introduced as possible surrogates for speech quality. One caveat is that several viable candidates require clean references (labels) for evaluation, which are not always available in real world tasks. Among several attempts to avoid clean label requirement, deep learning is one strong candidate for such tasks, due to the complex nonlinear functionality. Indeed, several neural network based approaches have been proposed to estimate speech quality~\cite{lo2019mosnet, zezario2022deep, mittag2021nisqa, reddy2022dnsmos,avila2019non}.

While neural network models provide a simple solution to get rid of clean reference requirement with seemly satisfying results, stability and consistency across different data are not ensured. In fact, it has been reported in several areas that certain imperceptible perturbations on input data can drastically alter network output, so that the prediction ability is heavily questioned. Such phenomena is usually caused by the so-called \emph{adversarial examples}, which prevail in both image and audio domain when using neural networks. Previous literature regarding audio adversarial examples was mainly focused on Automatic Speech Recognition (ASR) systems~\cite{carlini2018audio, qin2019imperceptible, yakura2018robust}. As the network-based speech quality prediction has gradually become a  trend and derived numerous downstream applications, it is important to carefully examine their prediction behavior in adversarial settings.

 


\textbf{Our contribution.} This work utilizes one well-known DNSMOS p.835 (non-intrusive quality) predictor to demonstrate that a deep-learning based model can be vulnerable to targeted adversarial attack. Our contribution contains two parts: (1) an approach to generate adversarial audio samples against the DNSMOS network is presented, while the adversarial attack is hardly noticeable to human ears, where the imperceptibility is supported by a human test. (2) we show that although the adversarial attack exposes the weakness of deep speech quality predictors, it can be used for model enhancement. Our experiments confirmed that the robustness can be strengthened by adversarial training.

\section{Adversarial examples causing inconsistent evaluations}\label{sec:adversarial}

\subsection{Speech quality prediction under perturbation}

This study investigates how quality predictions can be affected by small perturbations on input speech. Consider a speech quality prediction network $f$. An \emph{adversarial example} $\wtilde{x}$ of $f$ is an input data similar to another sample $x$ under certain measurement, such that the prediction $f(\wtilde{x}) \neq f(x)$. Due to the desired property that $\wtilde{x}$ should be close to an input $x$, adversarial examples are naturally considered as perturbations of input samples. As such, a (small) \emph{adversarial perturbation} $\d $ can also be defined whenever $\wtilde{x} = x + \d$ forms an adversarial sample.



The general description for targeted adversarial examples can be formulated as an optimization problem. Given $\eps \in \mathbb{R} $, an input $x \in \mathbb{R}^T$ and a target $y \in \mathbb{R}^k$, consider: 
\begin{equation}\label{E: targeted adversarial}
\min_{\d \in \mathbb{R}^T} L_S \left( x + \d , x \right) +  c \cdot L_T ( x + \d, y) \,\, \text{s.t.} \,\, D(\d) < \epsilon.
\end{equation}
where $L_S: \mathbb{R}^T \times  \mathbb{R}^T \to \mathbb{R}$ is a real-valued function measuring the \emph{similarity} between $x$ and the perturbed output $x + \d$. $L_T: \mathbb{R}^T \times  \mathbb{R}^T \to \mathbb{R}$ estimates the \emph{target deviation} between output $f(x + \d)$ from target $y$ such that when $f(x + \d) \to y $ one has $L_T ( x + \d, y) \to 0$. A coefficient $c \in \mathbb{R}$ is included to balance the two terms. When $c$ is large, the optimization naturally emphasizes $L_T$ and vice versa. $D : \mathbb{R}^T \to \mathbb{R}$ is a distortion metric for perturbations, introduced to be a constraint within tolerance $\eps$ allowed in a task.

In this study, we let $\mathbb{R}^T$ be the space of speech signals and choose $D=dB$ to measure the audio distortion in decibels ($dB$), which describes the relative loudness of a perturbation $\d = (\d_1, \ldots, \d_T) \in \mathbb{R}^T$ with respect to an input $x = (x_1, \ldots, x_T) $:
\begin{equation}
 dB_x(\d) =20 \, \log_{10} \left( \max_{t \in [0, T]} | \d_t | / \max_{t \in [0, T]}  |x_t| \right). 
\end{equation}


To confine perturbation decibel $dB_x(\d) < \eps $, the perturbation form $\d_t = A \cdot \tanh(z_t)$ is chosen with $A > 0$, $t=1,\ldots, T$. Since $\tanh(z_t) \in (-1, 1)$ for any $z_t \in \mathbb{R}$, the perturbation amplitude is always bounded $| \d_t | < A $. 

To construct a function faithfully reflecting similarity of two audio signal $\wtilde{x}$ and $x$, we consider comparisons in Fourier (spectrum) space under $L^1$-norm. Therefore, we define
\begin{equation}\label{E: sim loss}
L_S(\wtilde{x}, x) = \| \mathcal{F}(\wtilde{x}) - \mathcal{F}(x) \|_1
\end{equation}
with $\mathcal{F}$ the Short-Time Fourier Transform (STFT) onto Fourier space. The target deviation is chosen as:
\begin{equation}\label{E: target loss}
L_T (x, y) = \| f(x) - y \|_1
\end{equation}
where $\|\cdot\|_1$ is the $L^1$-norm. With the similarity loss Eq.~(\ref{E: sim loss}) and target loss Eq.~(\ref{E: target loss}) defined, together we derive the formulation for our adversarial task from Eq.~(\ref{E: targeted adversarial}):
\begin{equation}\label{E:obj}
\begin{aligned}
\min_{\d \in \mathbb{R}^T} & \| \mathcal{F}(x + \d) - \mathcal{F}(x) \|_1 + c \cdot \| f(x + \d) - y \|_1 \\
&\text{such that} \quad dB_x (\d) < \eps,
\end{aligned}
\end{equation}
This formulation is subsequently implemented to conduct adversarial training with small amplitude $A$. 

\subsection{Adversarial training to improve robustness}
\label{sec: enhance}



Although adversarial samples seems destructive to speech quality networks, there are occasions that they can be constructive. Below we explore the viability of enhancing the robustness using adversarial noises. 

Given a quality predictor $f$ and an audio sample $x_i$ from a speech corpus $\{x_i \}_{i=1}^N$, we consider the score $y_i = f(x_i)$ predicted by $f$ as a label such that the data pairs $\mathcal{D} = \{(x_i, y_i)\}_{i=1}^N$ are formed. Subsequently, given a target $\wtilde{y}$, an adversarial perturbation $\d_i$ can be derived by Eq.~\ref{E:obj} associated to each $x_i$ to achieve $f(x_i + \d_i) = \wtilde{y_i}$. When $\wtilde{y}_i \neq y_i$, the network $f$ is considered attacked. Particularly, when $\| \wtilde{y}_i - y_i \| $ is large with tiny $\d_i$, the network prediction is considered unstable.

To enhance a predictor with such type of weakness, we make the network be aware of adversarial examples. That is we correct ``false'' prediction $\wtilde{y_i}$ with the regular $y_i$ and achieve the defense. By collecting all adversarial examples, we can teach (retrain) the predictor with these irregular data pairs, called an \emph{adversarial dataset} $\mathcal{AD} = \{(x_i + \d_i, y_i)\}_{i=1}^N$.

Our goal is to derive a robustness-improved model $g$ from $f$ by training on an adversarially-extended dataset $\mathcal{D} \cup \mathcal{AD}$, where $\mathcal{AD}$ in this case can be regarded as data augmentation to strengthen the network stability. Our loss function for training process is defined as follows:
\begin{equation}\label{E: loss}
\begin{aligned}
\mathcal{L} (g) =\sum_{i}\| g\left( x_i  \right) -  f(x_i) \|_2^2,  \quad (x_i\in  \mathcal{D} \cup \mathcal{AD}).
\end{aligned}
\end{equation}
As the training is operated on two datasets $\mathcal{D}$ and $\mathcal{AD}$, there are two types of losses involved:
\begin{equation}\label{E: stage2 losses}
\mathcal{L}_1 (g) = \| g\left( x_i + \d_{x_i} \right) -  f(x_i) \|_2^2 , \,\, \mathcal{L}_2 (g) = \| g(x_i) -  f(x_i) \|_2^2,
\end{equation}
where $\|\cdot\|_2$ is the $L^2$ norm. We note that $\mathcal{L}_1$ intends to correct the adversarial perturbations with regular labels, and $\mathcal{L}_2$ serves as a forgetting loss~\cite{8107520}, which prevents $g$ from forgetting old knowledge inherited from $f$. Ideally, a new model $g$ is free from adversarial attack so that a perturbed audio has very similar scores as the unattacked $g\left( x_i + \delta_{x_i} \right) \cong  f(x_i)$. In the meantime, any unperturbed audio should maintain same score as before $g(x_i) \cong f(x_i) $. 




Recruiting adversarial data into training has been recognized effective in defending adversarial attacks, and model robustness is indeed found improved~\cite{goodfellow2014explaining, kurakin2016adversarial, madry2017towards}. Different from previous works where most of demonstrations were in image domain, this work devotes to speech quality assessment and intends to confirm the viability of adversarial training on speech quality models.

It should be noted that if a speech corpus has quality labels obtainable, one can always replace the surrogate index $y_i= f(x_i)$ with real (better) labels. Due to the inaccessibility in many corpus, it is our proposal to adopt $y_i= f(x_i)$ instead, which is probably more useful in numerous occasions.

\section{Experiments}
\label{sec:exp}

The following experiments were conducted with the released \textbf{DNSMOS P.835} CNN-based model, which predicts 3 subjective scores of noisy-clips based on ITU-T P.835, speech quality (SIG), background noise quality (BAK), and overall audio quality (OVRL)~\cite{dubey2022icassp}. The codes of this work shall be released upon acceptance of the manuscript.




\subsection{Adversarial examples on quality prediction model }\label{sec: ADV samples}

\subsubsection{Datasets}

\textbf{DNS-2020} is the dataset from 2020 DNS Challenge~\cite{reddy2020interspeech}, containing a noise set of 65,000 clips and 150 classes, selected from Audioset and Freesound. The clean speech data has 500 hours from 60,000 clips, obtained from the public audio books dataset named \emph{Librivox}. We adopted the resulting training dataset of 150 audio classes and 60,000 noisy clips.

\textbf{TIMIT} is a corpus frequently utilized in speech-related experiments. The speech data contains versatile acoustic-phonetic information including phonetically-compact sentences (SX) and phonetically-diverse sentences (SI), as well as regional diversity in dialects sentences (SA). A suggested \emph{core test} set consists of 192 sentences from 24 speakers, where 15 speakers were randomly selected in a balance manner to conduct the adversarial experiments.



\subsubsection{Experimental setting}


Under the pretrained weights of the DNSMOS network $f$, an adversarial perturbation $\d_{x, \wtilde{y}}$ was sought to attain a desired target (MOS) score $\wtilde{y}$ from input $x$ using optimization Eq.~(\ref{E:obj}). 


The STFT transformation $\mathcal{F}$ to measure $L_1$-similarity in Eq.~(\ref{E:obj}) had 512 Fourier basis ($n_{\text{FFT}} = 512$) under Hann window length 512 and hop size 128, resulting in 257 STFT dimensions denoting as frequency bins. The parameter $c=10$ was used in implementations. Input audio magnitudes were normalized, and the perturbations were set in the form $\d = 0.03 \cdot \tanh{z}$ so that the resulting $dB_x(\d) < -30 dB$.



We note that a target $\widetilde{y} = (\widetilde{y}_1, \widetilde{y}_2, \widetilde{y}_3) =(\texttt{SIG}, \texttt{BAK}, \texttt{OVRL})$ can have arbitrary subscore $\widetilde{y}_i 
\in [1,5]$. In our case, we intentionally consider utterly different scores to see interesting results. Particularly, we let $\widetilde{y}=(\widetilde{y}_1, \widetilde{y}_2, \widetilde{y}_3)$ to alter from the original prediction $y= (y_1, y_2, y_3)$ as follows:
\begin{equation}\label{E: target strategy}
\widetilde{y}_i = 
 \begin{cases}
              5 \qquad \text{if $y_i \in [1, 3]$,}\\
              1 \qquad \text{if $y_i \in (3, 5]$,}
  \end{cases} \,\, (i= 1, 2, 3)
\end{equation}
This relabelling strategy is interesting since a very clean speech with original high score $y=(5,5,5)$ is to be downgraded as $\wtilde{y}=(1,1,1)$ using adversarial perturbations. Contrarily, a very noisy audio with low predicted score $y=(1,1,1)$ is to be uplifted to $\wtilde{y}=(5,5,5)$. Another interesting case included is a mid-ranged score $y=(3.1, 2.8, 3.2)$ to be judged as $\wtilde{y}=(1, 5, 1)$, where the three MOS scores are torn to be contrasting.

\subsubsection{An example of results}

\begin{figure}[ht]
\vskip -0.0in
\begin{center}
\centerline{\includegraphics[width=\columnwidth]{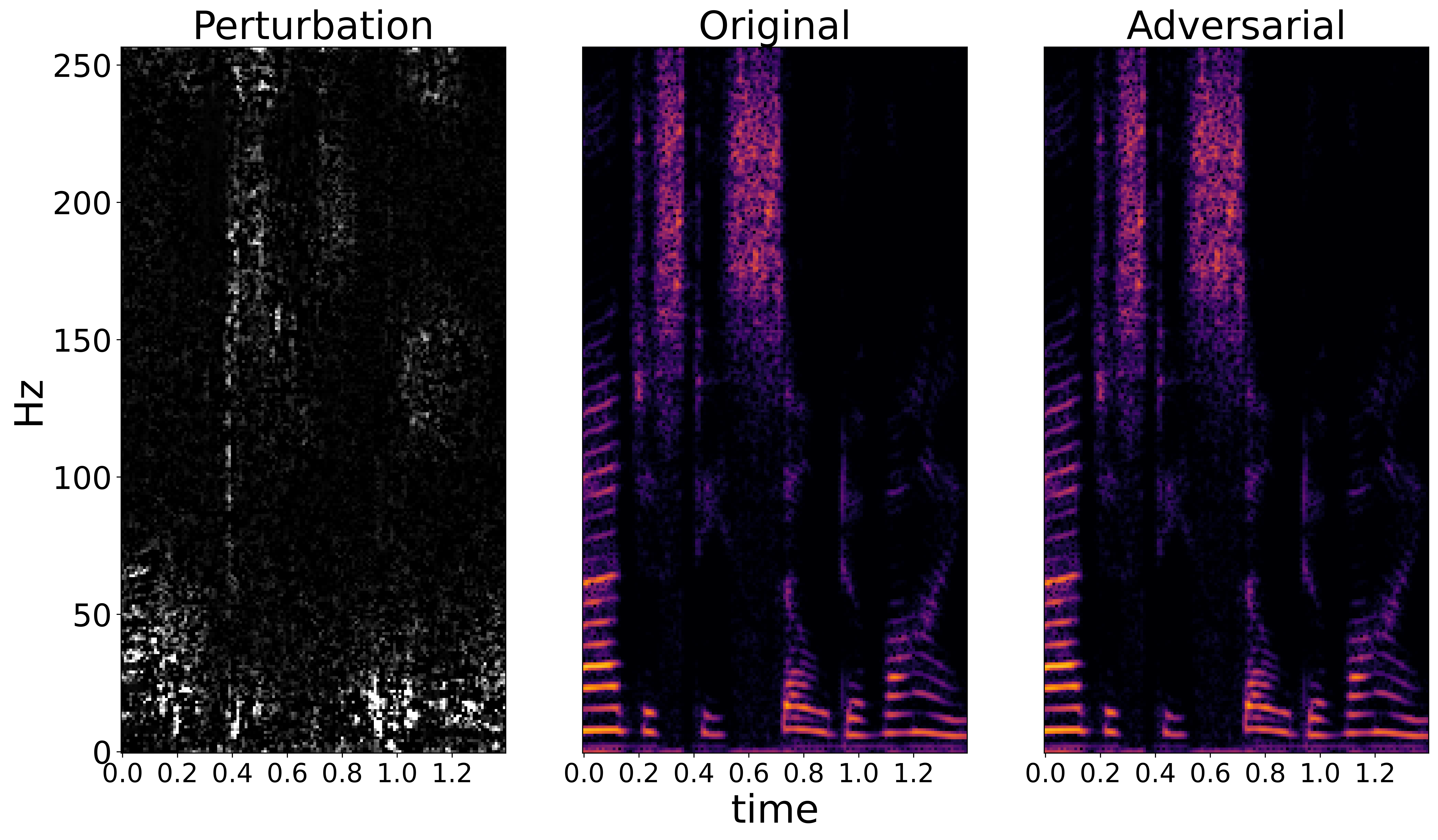}}
\caption{Visualization of an adversarial perturbation and its corresponding audio from DNS-2020. The perturbation is observed to conceal in utterances to sneakily alter scores.}
\label{fig: spec1}
\end{center}
\vskip -0.4in
\end{figure}

With limited space we demonstrate one adversarial example. In Fig.~\ref{fig: spec1}, an original audio (reader\_01326\_9\_7J3kchZ5UAg) from DNS-2020 and its adversarial correspondent are shown in spectrogram, where the prediction $y = (4.06, 4.16, 3.73)$ was downgraded to $\wtilde{y} = (0.99, 1.0, 1.0)$ by a small perturbation with small distortion $dB_x(\delta) = -36.01$ dB. This audio demonstration and more examples can be found at \url{https://hsinyilin19.github.io/}.

\subsection{Enhancing robustness by adversarial training}\label{subsec: enhance robust}


\subsubsection{Dataset}

\textbf{VCTK-DEMAND} is a noisy speech corpus premixed by the Voice-Bank (VCTK)~\cite{veaux2013voice} with real-world noises DEMAND database~\cite{thiemann2013diverse}. VCTK-DEMAND~\cite{valentinibotinhao16_interspeech} has 11,572 training samples and 824 testing samples composed by 28 speakers at 48 $k$Hz sampling rate. The VCTK-DEMAND corpus was used to conduct this enhancing experiment owing to the suitable data size for reasonable adversarial training time.


\subsubsection{Experimental setting}





In this experiment, adversarial perturbations were generated for each audio sample in both training and testing dataset of VCTK-DEMAND using Eq.~(\ref{E:obj}), (\ref{E: target strategy}). Consequently, the entire training set $\mathcal{D} = \{(x_i, f(x_i))\}_{i=1}^N$ yielded a corresponding adversarial set $\mathcal{AD} = \{(x_i + \d_{x_i, \wtilde{y}_i}, f(x_i))\}_{i=1}^N$ with $N=11,572$. A new network $g$ was trained by joint data $\mathcal{D} \cup \mathcal{AD}$ with initial weights from $f$. The loss function was $\mathcal{L}$ in Eq.~(\ref{E: loss}) and model $f$ was held fixed during the training process.




After training, the test set and its adversarial perturbations were used to verify the robustness of $g$. A model $g$ with output $(g_1,g_2,g_3)$ to claim an enhanced robustness should have the following property,
\begin{equation}\label{E: robust criteria}
| g_j( x_i + \d_{x_i} ) - f_j( x_i) | < | f_j( x_i + \d_{x_i} ) - f_j( x_i) |
\end{equation}
for any audio $x_i$ along with a perturbation $\d_{x_i}$ where $(f_1,f_2,f_3)$ are original predictions by $f$. Inequality~(\ref{E: robust criteria}) simply checks whether $g$ can better sustain adversarial perturbations than the original $f$ in recovering unperturbed score. For convenience, we denote the following errors:
\begin{equation}
\begin{aligned}
E_{g_j} &= \frac{1}{N}\sum_{i=1}^N \big| g_j( x_i + \d_{x_i} ) - f_j( x_i) \big| \\
E_{f_j} &= \frac{1}{N}\sum_{i=1}^N \big| f_j( x_i + \d_{x_i} ) - f_j( x_i) \big|\\
F_{g_j} &= \frac{1}{N}\sum_{i=1}^N \big| g_j( x_i  ) - f_j( x_i) \big|
\end{aligned}
\end{equation}
where $E_{f_j}$ computes the prediction deviation of $f$ and $F_{g_j}$ denotes the forgetting rate to check how much knowledge of $f$ is preserved in $g$.




\subsubsection{Enhancing results}
\begin{figure}[ht]
\vskip -0.0in
\begin{center}
\centerline{\includegraphics[width=0.8\columnwidth]{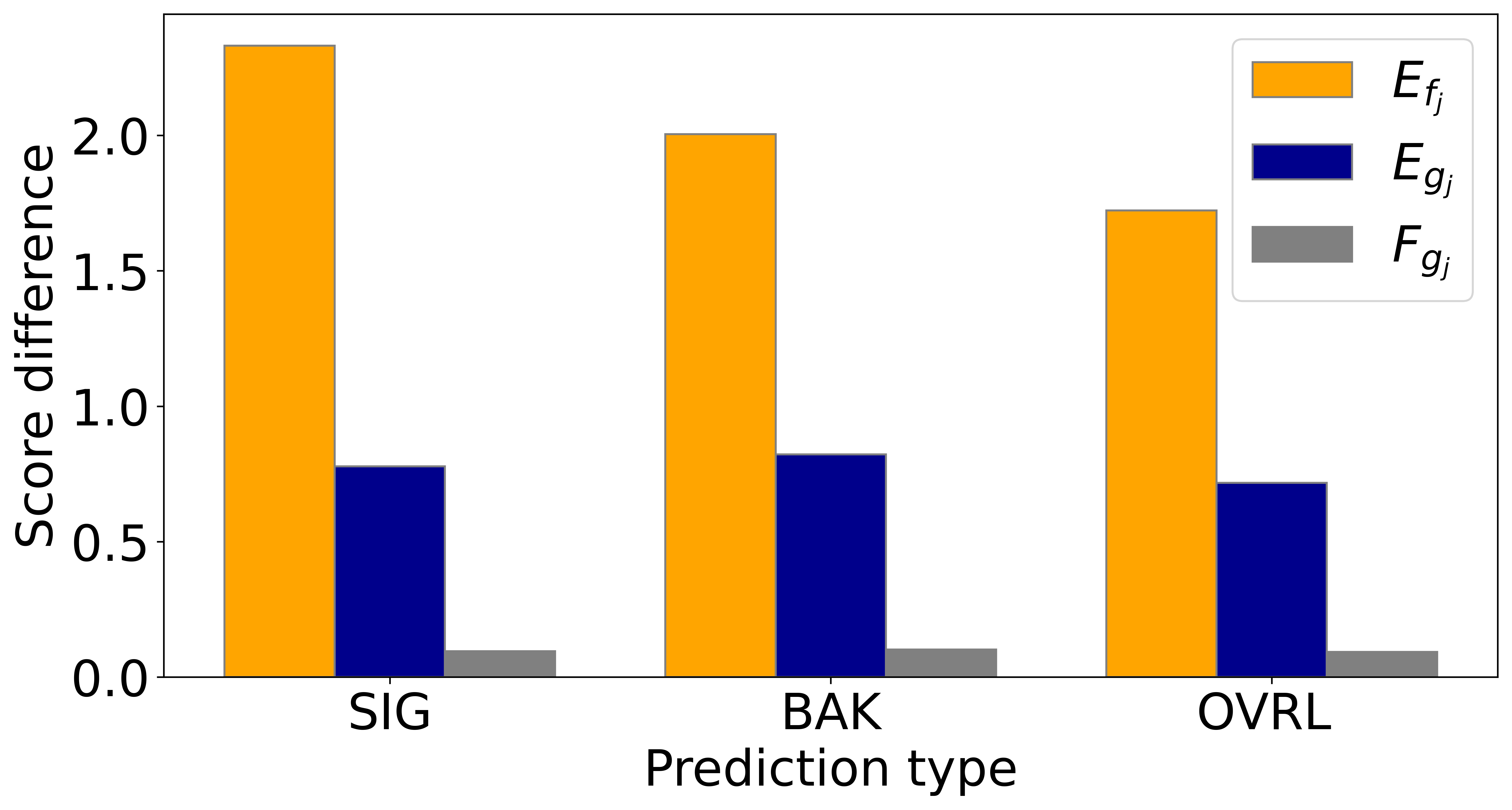}}
\caption{The score differences before/after adversarial training.}
\label{fig: enhance results}
\end{center}
\vskip -0.3in
\end{figure}

Fig.~\ref{fig: enhance results} shows the prediction deviation of $f$ and $g$. For $j=1,2,3$ (SIG, BAK, OVRL), the new deviation $E_{g_j}$ was observed to largely reduced down to less than half of the original deviation $E_{f_j}$. This clearly indicates that $g$ obtained better defense against \emph{unseen} adversarial perturbations on test set. In the mean time, small $F_{g_j}$ addresses that predictions of $g$ concurred with those of $f$ on the unattacked test audio. As such, the robustness of $g$ was indeed improved from $f$ to conclude our experiment.



\subsection{Human Imperceptibility Evaluation}

Having constructed numerous adversarial samples for the three datasets following procedure in Sec.~\ref{sec: ADV samples}, human evaluations were conducted to verify their imperceptibility.

In this evaluation, 35 participants were given 30 pairs of audio samples, asked to identify whether any difference might exist within each pair. The 30 pairs were composed of 10 randomly chosen pairs from each of the three datasets: DNS-2020, TIMIT, and VCTK-DEMAND. Among 10 pairs from each dataset, 7 pairs were adversarial; the other 3 were identically unperturbed ones. The participants were instructed to carefully answer either ``\textbf{A:} \emph{this pair is identical}'' or ``\textbf{B:} \emph{this pair has difference}''. The participants did not have time limit and the audio can be repeated until their answers were final.


\subsubsection{Results}
\begin{figure}[ht]
\vskip -0.0in
\begin{center}
\centerline{\includegraphics[width=\columnwidth]{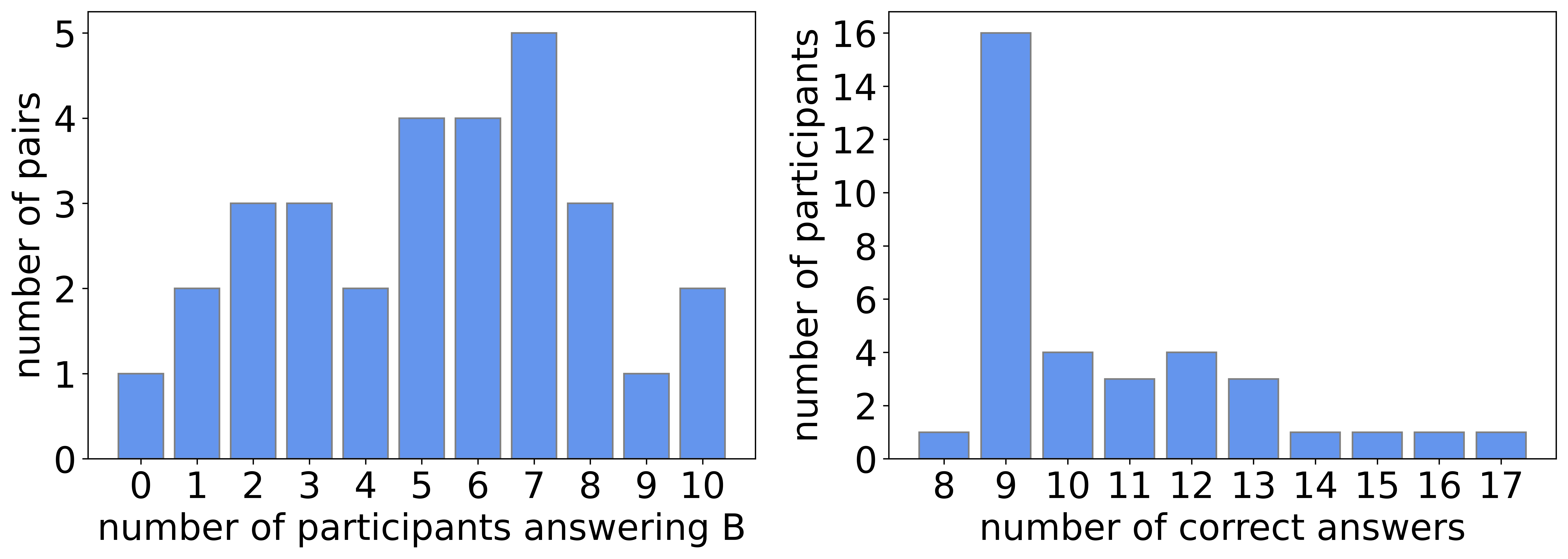}}
\caption{{[Left]} the figure shows the number of pairs in terms of the numbers of B received, \textit{e.g.} the $3^{rd}$ bar indicates there are 3 pairs with exactly 2 participants answered \emph{B}. {[Right]} it shows the number of pairs in terms of the numbers of correct answers in each questionnaire. \textit{e.g.}, the last bar shows there is only one participant who returned 17 correct answers. }
\label{fig: human}
\end{center}
\vskip -0.4in
\end{figure}

The results (Fig.~\ref{fig: human}) were examined in two perspectives. First, we counted each audio pair and it showed there were at most 10 (out of 35) persons chose ``B''. Moreover, there were only 2 (of 30 pairs) that received 10 B's from participants. Among the 2 pairs, one was adversarial; the other was in fact an identical. In brief, no matter a pair is identical or adversarial, there is always more than 71.43\% of the participants believing they are identical.

Secondly, we conducted statistical hypothesis testing for each participant. Let the null hypothesis be ``\emph{the participant cannot tell the difference between identical and adversarial pairs, and so he/she was guessing}''. Under the hypothesis, the $z$-score $ = 2 (x - 15) / \sqrt{30},$ where $x$ represents how many correct answers out of 30 questions each participant returned. After counting, we summarize that the all the participants returned between 8 and 17 correct answers. This implies that the $z$-score $\leq 0.73$ for all participants, equivalent to a one-tailed $p$-value $=23.27\%$. As the resulting $p$-values for all participants are fairly large, we conclude that it is very likely that all participants \emph{cannot} tell the difference between identical and adversarial pairs, and thus the adversarial perturbations are \emph{likely imperceptible} under the statistical sense.

\section{Conclusions}
\label{sec:conclusions}


In this work, we show that deep learning based speech quality predictors may be unstable under adversarial perturbations, where we use DNSMOS P.835 to demonstrate such vulnerability exists and may result in unreasonable quality ratings. This further suggests that a network predictor to apply to downstream tasks should be carefully examined. The study contributes to this matter further as we explore the possibility to strengthen network robustness by adversarial training. Our preliminary result on DNSMOS verifies the approach is effective for speech quality predictors and promising for future investigation.

\end{document}